\def\year{2019}\relax
\documentclass[letterpaper]{article} 
\usepackage{aaai19}  
\usepackage{times}  
\usepackage{helvet} 
\usepackage{courier}  
\usepackage[hyphens]{url}  
\usepackage{graphicx} 
\usepackage{comment}
\def\UrlFont{\rm}  
\usepackage{graphicx}  
\title{Experiences with Improving the Transparency of AI Models and Services}
\author{
Michael Hind, Stephanie Houde, Jacquelyn Martino, Aleksandra Mojsilovic,\\[.5ex]
{\Large \bf David Piorkowski, John Richards, and Kush R.\ Varshney}\\
IBM Research
}

\begin{document}
\raggedbottom

\maketitle

\begin{abstract}
AI models and services are used in a growing number of high-stakes areas, resulting in a need for increased transparency. Consistent with this, several proposals for higher quality and more consistent documentation of AI data, models, and systems have emerged. Little is known, however, about the needs of those who would produce or consume these new forms of documentation. Through semi-structured developer interviews, and two document creation exercises, we have assembled a clearer picture of these needs and the various challenges faced in creating accurate and useful AI documentation. Based on the observations from this work, supplemented by feedback received during multiple design explorations and stakeholder conversations, we make recommendations for easing the collection and flexible presentation of AI facts to promote transparency.
\end{abstract}

\section{Introduction}
AI models and services are being used in a growing number of high-stakes areas such as financial risk assessment \cite{credit-2018}, medical diagnosis and treatment planning \cite{health-care-2019}, hiring and promotion decisions \cite{hiring-2017}, social services eligibility determination \cite{social-services-2019}, predictive policing \cite{ensign2017runaway}, and probation and sentencing recommendations \cite{propublica}.

While most models are created for bespoke purposes, some are also being packaged in model catalogs for use by others. For many models there will be risk, compliance, and/or regulatory needs for information covering the nature and intended uses of the model, its overall accuracy, its ability to explain particular decisions, its fairness with respect to protected classes, and at least high-level information about the provenance of training data and assurances that suitable privacy protections have been maintained. In reviewing models within a catalog for suitability in a particular application context, there may be an additional need to easily compare multiple candidates.

Recent work has outlined the need for increased transparency in AI for data sets \cite{gebru-2018,data-statements,HollandHNJC2018}, models \cite{model-cards}, and services ~\cite{factsheets-2019}. Proposals in support of ethical and trusted AI are also emerging \cite{EuropeanCommission2019,partnershipOnAI2019,ieee-2017}. While details differ, all are driving towards a common set of attributes that capture essential ``facts" about a model. We are not yet aware of developers adopting these ideas for regular use. Neither are we aware of published work describing developers' needs or the difficulties they face when producing or consuming AI documentation.

In this paper we discuss formative research with developers and other stakeholders to better understand their documentation needs. We also report on a study in which developers in several application areas created AI documentation in the form of a {\em FactSheet}.

A FactSheet, as proposed by \cite{factsheets-2019}, is a collection of relevant information about an AI model or service that is created during the machine learning life cycle. It includes information from the business owner (e.g., intended use and business justification), from the data gathering/feature selection/data cleaning phase (e.g.,  data sets, features used or created, cleaning operations), from the model training phase (e.g., bias, robustness, and explainability information), and from the model validation and deployment phase (e.g., key performance indicators).  A FactSheet is associated with a model (or service) and is meant to be write once, i.e., an update to a model would trigger a new FactSheet for the updated model. FactSheets can be consumed by any role in the ML life cycle to confirm process governance adherence or model performance, or by the ultimate users of a model to provide increased transparency. Of course, the diversity of model types, and the range of possible application domains, makes the specification of a common FactSheet schema difficult. We hope the work reported here provides useful guidance going forward.

The contributions of this paper include
\begin{itemize}
\item summaries of semi-structured interviews with AI developers on their documentation needs and practices, 
\item observations on documentation requirements and difficulties of AI developers in creating FactSheets, 
\item additional requirements from feedback on prototype FactSheet designs and unstructured interviews with stakeholders involved throughout the AI life cycle, and 
\item recommendations for supporting mechanisms and new research to improve FactSheet creation.
\end{itemize}

The remainder of this paper is organized as follows: Section 2 discusses related work on the creation of software documentation. Section 3 outlines how we engaged with AI model and service owners to understand potential FactSheet content in a variety of application domains. Section 4 presents our primary observations about the creation of model facts, focusing on both general requirements and particular challenges AI developers faced. Section 5 recommends approaches to addressing these needs and challenges, and Section 6 offers concluding thoughts.

\section{Related Work}

The challenge of creating useful and usable documentation is not new. Software engineering research has identified quality issues in existing documentation for conventional systems \cite{garousi2013evaluating,robillard2011field,sohan2017study} and identified problems such as missing rationales for design decisions, too few examples to understand how to use a module or package, lack of overviews to illustrate how a system's component parts work as a whole, and insufficient guidance on how to map usage scenarios to elements of an API. Exacerbating these problems is the fact that documentation tends to be costly to create and maintain, and is often left as a low-priority item during software development \cite{uddin2015api}. Perhaps unsurprisingly, developers may ignore documentation altogether and prefer to read source code \cite{maalej2014comprehension} or inspect the outputs returned to various inputs, as they view these techniques to be a more reliable reflection of reality.

Despite these difficulties, developers do consider proper documentation to be a critical part of good software engineering \cite{robillard2011field}. Research has investigated the needs and practices of documentation consumers \cite{forward2002relevance,maalej2014comprehension,meng2018application} examining areas such as how to support problem solving when the required information is scattered across multiple sources or when the documentation is simply too general to provide specific answers to developers' questions. Given these challenges, researchers have advocated for more automated approaches \cite{robillard2017demand} and have explored systems for automatic documentation for various use cases \cite{mcburney2014automatic,subramanian2014live,sohan2015spyrest,zhang2011automated}.

Given the reliance of AI models on training data, and the often probabilistic behavior of AI systems with respect to test data, we believe that AI documentation will include features not found in documentation of general software. The emergence of data sheets \cite{gebru-2018,data-statements,HollandHNJC2018}, model cards \cite{model-cards}, and FactSheets~\cite{factsheets-2019} attest to these differing needs. We turn now to how we explored the requirements and possible benefits of AI FactSheets and the potential difficulties faced by developers in creating and consuming them.

\section{Methodology}

Our formative research into potential FactSheet uses, requirements, and challenges consisted of two primary threads. First, we conducted semi-structured interviews with six data scientists to investigate potential FactSheet use cases. Second, we led two FactSheet creation sessions with AI developers to understand their views on what content to include in FactSheets and the difficulties they faced in actually creating that content. In this paper we report on the results from this work, along with additional observations from multiple design explorations and unstructured interviews with over a dozen stakeholders involved in various aspects of AI model or system requirements, development, testing, deployment, and monitoring.

\subsection{Data Scientist Interviews}
We conducted semi-structured interviews focused on current AI documentation practices and potential FactSheet use cases. We recruited six participants (one female and five males), half from within our research and development organization and half from outside. Participants' experience in their current role ranged from four months to twenty years. Their background and experience included applied mathematics, data analytics, computer science, cognitive science, engineering and business analytics, and machine learning.  Participants suggested several use cases where FactSheets could be valuable, summarized in Table \ref{tbl:use-cases}.
 
\begin{table}[htb]
\centering
\scriptsize
    \begin{tabular}{|p{0.1in}|p{1in}|p{1.65in}|}
        \hline
        P\# & Role                      & Use Case Summary \\ 
        \hline
        1   & Data Scientist            & To understand models inherited from an outside source. To compare models. \\ 
        \hline
        2   & Research Scientist        & To test models when original data is not available. \\
        \hline
        3   & Machine Learning Intern   & To review work with supervising data scientist. To compare models. \\ 
        \hline
        4   & Data Scientist            & To generate reports for business manager. To compare models. \\ 
        \hline
        5   & Data Scientist            & To generate well-structured internal documentation. \\ 
        \hline
        6   & Data Scientist            & To solve problems that surface in testing. \\ 
        \hline
    \end{tabular}
\caption{Interview participants' use cases.}
\label{tbl:use-cases}
\end{table}
 
Participants P1 and P2 talked about how FactSheets could be used to improve model comprehension. For P1, this was in the context of trying to understand a model developed by others that had low accuracy on new test data. For P2, this was a situation where they were unable to access the original data set, but wanted a mechanism for at least comparing distributions of the new data set to the original. FactSheets could help in these situations by providing key details about model creation decisions and information about training and test data that would otherwise be unavailable or difficult to uncover.

P3 and P4 reported the need to facilitate discussions about models with others. P3 was required to provide reports explaining why they were using a particular model over others. To do so, they generated lengthy reports highlighting key characteristics or metrics across multiple models. Similarly, P4 wanted to provide details of model performance to their business manager who did not have a background in data science or statistics. FactSheets could help facilitate discussions between team members by providing one report view for those who are more statistically inclined and another for those who are not.

The final two participants, P5 and P6, envisioned use cases where FactSheets were more closely integrated within their development environments. P5 wanted to provide well-structured internal documentation to support reproduction of results by other researchers. They discussed a desire to provide a summary of key facts about a model inline with the Jupyter notebook used to build the model. P6 envisioned a closer integration of FactSheet to the testing tools they were using and suggested including remediation recommendations if problems were found.

Taken together, these discussions helped refine FactSheet requirements. They led to identifying information that should be included in a FactSheet (P1 and P2), suggested how FactSheets can be used to overcome communication challenges (P3 and P4), and raised potential opportunities for further integration into AI engineering tools and practices (P5 and P6).

\subsection{FactSheet Creation Session 1}
To gather more detailed FactSheet requirements, and to understand the difficulties developers might face in generating useful facts, we recruited nine AI model or service developers from within our larger organization. Their models varied both in type and in application domain.

Each developer was first introduced to the idea of FactSheets as the primary source of information about a model. Each was then given a sample FactSheet consisting of 39 questions and answers based on ~\cite{factsheets-2019} and shown in Table \ref{tbl:39qs}. They were then asked to create a FactSheet to document their model, the stated intent of this documentation to be for use in a model marketplace where users could search for, compare, and select models appropriate for their tasks. Developers were encouraged to add or remove any facts as they deemed necessary.

\begin{table*}[t]
\centering
\scriptsize
    \begin{tabular}{|p{2.1in}|p{2.1in}|p{2.1in}|}
        \hline
        \begin{enumerate}
            \itemsep 0em
            \item What is this service about? 
            \item Describe the outputs of the service. 
            \item What algorithms or techniques does this service implement? 
            \item What are the characteristics of the development team? 
            \item Have you updated this FactSheet before? 
            \item What is the intended use of the service output? 
            \item What are the key procedures followed while using the service? 
            \item What are the domains and applications the service was tested on or used for? 
            \item How is the service being used by your customers or users? 
            \item List applications that the service has been used for in the past. 
            \item Which data sets was the service tested on? (e.g., links to data sets that were used for testing, along with corresponding data sheets). 
            \item Describe the testing methodology. 
            \item Describe the test results. 
            \item Is there a way to verify the performance metrics (e.g., via a service API )? 
            \item In addition to the service provider, was this service tested by any third party? 
        \end{enumerate}
        &
        \begin{enumerate}
            \itemsep 0em
            \setcounter{enumi}{15}
            \item Are you aware of possible examples of bias, ethical issues, or other safety risks as a result of using the service?  
            \item Do you use data from or make inferences about individuals or groups of individuals. Have you obtained their consent?  
            \item Are the service outputs explainable and/or interpretable? 
            \item For each data set used by the service: Was the data set checked for bias? What efforts were made to ensure that it is fair and representative?  
            \item Does the service implement and perform any bias detection and remediation? 
            \item What is the expected performance on unseen data or data with different distributions? 
            \item Does your system make updates to its behavior based on newly ingested data? 
            \item How is the service tested and monitored for model or performance drift over time? 
            \item How can the service be checked for correct, expected output when new data is added? 
            \item Does the service allow for checking for differences between training and usage data? 
            \item Do you test the service periodically?
            \item How could this service be attacked or abused? Please describe. 
        \end{enumerate}
        &
        \begin{enumerate}
            \itemsep 0em
            \setcounter{enumi}{27}
            \item List applications or scenarios for which the service is not suitable. 
            \item How are you securing user or usage data? 
            \item Was the service checked for robustness against adversarial attacks? 
            \item What is the plan to handle any potential security breaches? 
            \item Does the service provide an as-is/canned model? Which data sets was the service trained on? 
            \item For each data set: Are the training data sets publicly available? 
            \item For each data set: Does the data set have a data sheet or data statement? 
            \item Did the service require any transformation of the data in addition to those provided in the data sheet? 
            \item Do you use synthetic data? 
            \item How were the models trained? 
            \item When were the models last updated? 
            \item Did you use any prior knowledge or re-weight the data in any way before training?  
        \end{enumerate}
        \\
        \hline
    \end{tabular}
\caption{Sample FactSheet questions. All models were deployed as services, making model and service interchangeable.}
\label{tbl:39qs}
\end{table*}

\subsection{FactSheet Creation Session 2}
For the second FactSheet creation session, we reengaged with six of the nine developers and provided ten high-level questions to be answered during one-hour co-development sessions with one of the authors.  The ten questions are presented in Table \ref{tbl:short-facts} and were selected as a commonly recurring subset from the first session. We asked participants to again fill out a FactSheet for a potential user of a model marketplace. After they completed this task, one of us walked them through their filled-out FactSheet to extract rationales for \emph{why} the information they provided was provided.

\begin{table}[htb]
\centering
\scriptsize
    \begin{tabular}{|p{1.4in}|p{1.4in}|}
        \hline
        \begin{enumerate}
            \itemsep 0em
            \item What is this model for?
            \item What domain was it designed for?
            \item Information about the training data (if appropriate)?
            \item Information about the model (if appropriate)?
            \item What are the model's inputs and outputs?
            \item What are the model's performance metrics? (Accuracy, Bias, Robustness, Domain Shift, Other metrics that you think are appropriate for this model)
        \end{enumerate}
        &
        \begin{enumerate}
            \itemsep 0em
            \setcounter{enumi}{6}
            \item Information about the test set?
            \item Can a user get an explanation of how your model makes it decisions?
            \item In what circumstances does the model do particularly well (within expected use cases of the model)? (e.g., inputs that work well)
            \item Based on your experience in what circumstances does the model perform poorly? (e.g. domain shift, specific kinds of input, observations from experience)
        \end{enumerate}
        \\ \hline
    \end{tabular}
\caption{Reduced ten-question FactSheet questions.}
\label{tbl:short-facts}
\end{table}

From these rationales we learned that developers wanted to include metrics and information about their models that conform to the conventions typical for their model type or application domain. For example, in a language translation model, the developer found it important to include a BLEU score ~\cite{bleu-2002}, which the language translation community uses to compare the efficacy of translation models. We also learned that developers' familiarity with a model often led them to omit information useful to someone less informed about the model type. Lastly, although model owners were appreciative of the reduced length of the ten-question FactSheet, they still found the task difficult and time-consuming.

\section{Findings}
In this section we report our major findings from  these  two  FactSheet  creation  exercises, and  the comments gathered during the interviews described above.

\subsection{Perceived FactSheet Value}
All but the most senior interviewee (with well-established work practices), and all the developers in our FactSheet creation sessions, viewed FactSheets as valuable. The idea of capturing key facts about an AI model or service in a form that is useful to a broad range of stakeholders was appealing. AI developers stated that FactSheets could provide useful guidance on how best to document their work. Those who have tried to consume models developed by others noted the importance of understanding how the models were structured, what data was used to train them, how features were engineered, and why a particular model or class of model was selected as fit for purpose. FactSheets could and should include these elements. 

This research, along with feedback from multiple design explorations and unstructured conversations with stakeholders involved in the creation, deployment, or monitoring of AI systems, have revealed that a substantial part of FactSheets' appeal is the belief that much of what is important about a model can be automatically captured. Some facts, such as typical accuracy measures, can be computed automatically on many platforms. Other facts, such as a model's intended use or the enumeration of situations for which a model may be inappropriate, cannot be automatically captured. For an extreme example of this, consider just a few of the questions in the checklist now being piloted in the EU \cite{EuropeanCommission2019}: ``Could the AI system affect human autonomy by interfering with the (end) user’s decision-making process in an unintended way?"; ``Did you take safeguards to prevent overconfidence in or over reliance on the AI system for work processes?"; ``Did you identify potential safety risks of (other) foreseeable uses of the technology, including accidental or malicious misuse? Is there a plan to mitigate or manage these risks?". These important questions can only be answered by knowledgeable humans. 

\subsection{Observed FactSheet Challenges}
What particular difficulties did developers face when trying to answer questions for a FactSheet? Some developers simply forgot important details such as how they transformed training data or which hyper parameter manipulations they explored along the path to a final model. While forgetting is not surprising, we were still impressed with just how quickly details slipped away. For example, in creating one of our FactSheets, recovering information about training data transformations required several hours over the course of two days. This was true even though the original data set was relatively small (25 columns by 1000 rows, reduced to 12 columns after transformation) and the data scientist who performed these transformations was still available. Eventually, they found the Python notebook that performed the data transformations and the history was reconstructed.

Another area of concern for our participants was documenting facts about a model that might reflect poorly on data provenance, testing protocols, or the possibility of various biases in either the training data or model output. With respect to bias we noted that most developers assumed it was not going to be an issue even if further probing by us revealed it to be a possibility. Additional unstructured interviews have confirmed that AI developers are quite unfamiliar with how bias might manifest in their work. As such it is not surprising that our developers simply asserted that a question about bias was not applicable rather than consider the various ways that bias could surface in different contexts. Even for developers familiar with bias metrics they might be inclined to merely state that there are no protected variables in the training data rather than look for features that may be correlated with protected variables or enumerate more subtle ways in which use could lead to biased outcomes.

Several participants noted that data or model details may be proprietary. It is often not clear where to draw the line between providing enough information for a model to be adopted while not revealing information that threatens competitive advantage. For an example of this problem see \cite{propublica}. 

The reason that FactSheets or related forms of documentation will be produced is so others can benefit from consuming them. In many cases, however, we observed that FactSheet producers had significant difficulties creating FactSheets for potential consumers, with the developers reporting that it was difficult to know what information was going to be needed. This may be especially true in the case of models packaged for reuse in shared catalogs. How is it possible, short of extensive experience in a domain, to anticipate the ways that models might be used, or misused? More generally, how is it possible to know whether a FactSheet will be viewed from the perspective of say a testing team as opposed to industry regulators? Each consumer will have different levels of understanding and will be performing quite different tasks. 

We have found two exceptions to this lack of clarity about documentation consumers, one in the area of human resource management and the other in financial services. In the case of human resources, we interviewed a team member overseeing the annual updating of retention and salary recommendation models that are reviewed and approved by a known set of stakeholders representing specific business units and geographies (geography being especially interesting insofar as labor laws vary considerably around the world). In the case of financial services we have interviewed domain experts who note that teams composed of people other than the original model developers are often specifically tasked with writing extensive model documentation (on the order of 100 pages) for review by known regulatory agencies with known assessment criteria. Beyond these cases, however, the general problem of understanding the highly variable capabilities and needs of different consumers will likely complicate the production of useful FactSheets.

Finally, many participants expressed a desire to compare two or more FactSheets. Common reasons included presenting several related models to highlight the superiority of a newly developed one or the need to select one model from among alternatives in a catalog. This highlights another difficulty for those creating FactSheets; there are many ways to meaningfully assess the fairness, robustness, and even the accuracy of models. Based on the emergence of standards in other domains, we believe this situation will improve over time as useful patterns emerge and are adopted (the probable time course varying by domain and model type). We also believe that tool kits of common metrics will eventually stabilize and be broadly used (see, for example \cite{aif360-2019}). In the meantime, we have found that AI developers desire guidance on how best to document even the most common facts about a model.

\section{Recommendations}

We have found a range of needs for those creating and consuming AI documentation. We have also found considerable diversity across domains and model types as to which facts in an AI FactSheet are likely to be useful. Despite this diversity we have learned enough to suggest some directions for future work that could lead to better, more consumable, FactSheets and more widespread FactSheet adoption.

\subsection{Fact Collection}
Nascent ``facts", in the form of discreet events, are already arising throughout the AI life cycle. Most of these potentially interesting facts go unnoticed. Others are noticed but quickly forgotten. A few are noticed and recorded (with varying fidelity) in unconnected systems that prevent coherent documentation from being produced reliably or efficiently. This is the current situation. If it is to change we must first acknowledge that facts will arise from the actions of different stakeholders (ranging from line of business managers, to data scientists, to risk and compliance officers) and that they will arise through the use of many different tools (from business modeling software, to Jupyter notebooks, to risk management and reporting systems). Consider a few examples. Intended use descriptions and required performance targets might be specified by a line of business owner. Data provenance and licensing details might be documented by a data steward. Accuracy, bias, and robustness metrics might be generated by a data scientist or quality assurance team. Some facts will be generated automatically as a side effect of a data transformation or a training cycle. Others will arise in discussions between high-level managers in operations meetings. Importantly, whatever we can do to make fact collection easy, either by making it more automatic or simply making it possible to write (and post) a bit of descriptive text in the moment of tool use, will decrease the incidence of forgetting the fact or forgetting important details about the fact.

It is certainly \textit{possible} to create a single integrated system including all the tools used throughout the AI life cycle, each tool equipped with a common mechanism to collect facts for later use. But it is likely that such a system would impose unacceptable constraints on the tools that organizations want to use and the way that organizations want to work. We believe a more realistic approach is to define an open API for registering models, for posting facts about them, and for retrieving them for monitoring and reporting. This API could define the end points for a pub-sub architecture, such as Pulsar \cite{pulsar2019}, enabling the creation of a FactSheet Repository for a diverse and essentially unbounded set of tools. 

\subsection{Fact Authoring}
Some important AI facts can be captured automatically. Others, perhaps those that have coalesced around a stable set of practices within a development community, can be readily created. Some facts, however, will require careful thought about what should be documented. Our developers found it difficult, for example, to specify the boundary conditions beyond which model use was inadvisable. In another example, developers were often unsure about the level of detail to include in descriptions of a model's structure. We have found through our interviews and FactSheet creation exercises that this sort of human fact ``authoring" is challenging and the quality of authored facts is quite variable. Both productivity and quality can be improved, however. For example, if user testing indicates that a particular FactSheet question is hard to understand, the question can be clarified through a cycle of user testing and refinement. Alternatively, if the question is understandable but answers are often incomplete or of poor quality, hints or examples of well-formed answers can be offered.

Even in our somewhat limited testing we have found that FactSheets will often include questions for which there is no relevant answer. ``N/A" may be a perfectly acceptable fact value in this case, but it should be applied thoughtfully. Some facts may be known but proprietary (and could be redacted for those with insufficient access rights). In other cases, model builders will assume that some kinds of facts are not applicable even if, on further reflection, they are. We have seen that model bias is often considered inapplicable. To get better answers in cases such as this we may need to invent elicitation techniques that go beyond mere form filling. For example, if a question is frequently marked as not applicable a wizard can be created to walk the user through a process that will lead, in the end, to a suitable answer. 

We have also noted that different ways of recording the same information can make model comparisons difficult. Just as answer quality can be improved through hints and examples, so too can answer consistency. More generally, open fact schemata can be created, perhaps even standardized, to promote greater consistency within particular domains or industries. On a smaller scale, an organization that is finding unnecessary variation in test protocols or choice of performance metrics could configure its AI pipelines to provide only a restricted set of options. Finally, we consider it worth exploring whether some form of FactSheet Policy, registered with the FactSheet Repository, could be used to control which facts are collected during AI development.

\subsection{Fact Retrieval and Reporting}
At some point, facts about a model that are collected automatically or created relatively straightforwardly, along with facts that are more laboriously authored, will generally (but not necessarily) be assembled and rendered into coherent documents. Of course, not all facts about a model need to be seen by all stakeholders. And in some situations, not all facts about a model \textit{should} be seen by all stakeholders. The above-mentioned redaction of proprietary facts for those without adequate permissions is an example of this. It follows that facts may need to be tagged in a way that permits selection of only a subset of them, based on stakeholder access rights. Providing this sort of control, assuming a good set of tags can be created, will need to be part of a complete FactSheet solution.

Beyond the need for facts to be excluded because of limited access rights, there is a need for facts to be assembled in different ways when they are rendered as FactSheets for different stakeholders. A test team, for example, may not need to see the details of model training or the business key performance indicators that will be monitored once a model goes into production. A risk and compliance officer or an external regulator may not need to see a model's internal structure but may need to certify that certain tests were successfully performed or that data privacy standards were adhered to. One way to manage this diversity of views is through the use of what we could call {\em FactSheet Templates}. FactSheet templates could be created through a template builder drawing on an inventory of all fact types that an organization wanted to collect and report. A particular template would determine the content and layout of the associated FactSheet instance generated for a particular stakeholder class. To avoid the creation of new mechanisms, templates could refer to one or more CSS files to control the presentation. Like FactSheet Policies, FactSheet Templates could be registered in the FactSheet Repository for use as needed.

In addition to excluding or including particular facts in a generated report, we have found at least one case where different stakeholders needed to see the \textit{same} fact at different levels of detail. Not surprisingly, the prospect of creating a set of different fact forms for the same fact was viewed as unattractive. To support this case, information within a fact could be more finely structured such that portions of it were individually addressable. Alternatively, facts consisting of lengthy unstructured text might be automatically summarized for those needing only an overview. Future research might profitably focus on meeting this need.

\section{Concluding Thoughts}

Accurate and understandable facts about a model throughout its full life cycle --- from requirements specification, to data curation and feature engineering, to training and testing, to deployment and monitoring --- will provide a range of benefits, some of which we can only speculate about now. For most models, there are currently no such well-assembled facts.

This is not just a pain point for developers and a cost driver for organizations. The absence of useful information diminishes the perceived trustworthiness of the models we create. When important model facts are easily collected, authored, and reported, trust and responsible AI use will grow.

The focus of this work is on the requirements and challenges of
creating FactSheets, a necessary step to improve the governance and
transparency of AI.  Another important dimension is the level of trust
in the facts themselves.  Is it sufficient to have an enterprise self-report
their facts or do standards bodies or third-party
certification agencies conduct or validate this reporting?  The EU~\cite{EuropeanCommission2019} is looking at this
issue and there is legislation beginning to emerge at the state level in the US~\cite{NJ-2019}.

We have discussed the reported experiences of developers in creating and consuming existing AI documentation. We have also worked with developers as they have tried to create the form of documentation proposed as FactSheets. The problems they faced, and feedback from multiple design explorations and unstructured interviews with various stakeholders, have led to a series of recommendations for improving system support for fact collection, human fact authoring, and flexible fact reporting. Future explorations of these ideas may lead to a new, open ecosystem improving our collective understanding of the AI models, services, and systems we create.

\pdfinfo{
/Title (Experiences with Improving the Transparency of AI Models and Services)
/Author (Blind)
/Keywords (Input your keywords in this optional area)
}

\bibliography{refs}
\bibliographystyle{aaai}

\end{document}